\documentclass[conference]{IEEEtran}
\IEEEoverridecommandlockouts
\usepackage{cite}
\usepackage{hyperref}
\usepackage{amsmath,amssymb,amsfonts}
\usepackage{amsmath}
\DeclareMathOperator*{\argmax}{argmax}
\usepackage{graphicx}
\usepackage{textcomp}
\usepackage{xcolor}
\usepackage{graphicx}
\usepackage{float} 
\usepackage{subfigure}
\usepackage{amsmath}
\usepackage{amsfonts,amssymb}
\usepackage{mathrsfs}
\usepackage{mathtools}
\usepackage{algorithm}
\usepackage{algorithmicx}
\usepackage{algpseudocode}
\bibliography{IEEEabrv}
\def\BibTeX{{\rm B\kern-.05em{\sc i\kern-.025em b}\kern-.08em
    T\kern-.1667em\lower.7ex\hbox{E}\kern-.125emX}}

\begin{document}

\title{Optimal Transmit Antenna Selection Algorithm in Massive MIMOME Channels\\
\thanks{This paper will be presented at IEEE Wireless Communications and Networking Conference (WCNC), 2019. And thanks for the support from my coauthor Mr. Zeliang Ou and my supervisor Mr. Hongwen Yang.}
}

\author{\IEEEauthorblockN{Chongjun Ouyang, Zeliang Ou, Lu Zhang and Hongwen Yang}
\IEEEauthorblockA{\textit{Wireless Theories and Technologies Lab} \\
\textit{Beijing University of Posts and Telecommunications}\\
Beijing 100876, China \\
\{DragonAim, ouzeliang, zhangl\_96, yanghong\}@bupt.edu.cn}
}

\maketitle

\begin{abstract}
This paper studies the transmit antenna selection in massive multiple-input multiple-output (MIMO) wiretap channels, also termed as multiple-input multiple-output multiple-eavesdropper (MIMOME) channels. The transmitter, equipped with a large-scale antenna array whose size is much larger than that of the legitimate receiver and eavesdropper, selects a subset of antennas to transmit messages. A branch-and-bound (BAB) search based algorithm for antenna selection in independent and identical distributed Rayleigh flat fading channel is proposed to maximize the secrecy capacity between the transmitter and the legitimate receiver when the transmit power is equally allocated into the selected antennas. Furthermore, the proposed algorithm is separately applied to two scenarios which is based on whether
the channel side information of the eavesdropper (CSIE) is available at the transmitter. Simulation results show that the proposed algorithm has the same performance as the exhaustive search under both scenarios but with much lower complexity.
\end{abstract}


\section{Introduction}
\label{sction1}
Data flux in wireless networks has experienced an explosive growth with the sharp increment of the amount of smart devices. With the rapid growing demand for transmission rate, the significance of transmission security and reliability has become increasingly prominent. In this respect, physical layer (PHY) security {\cite{b1}} has gained pivotal attention in recent years for its remarkable performance in information security enhancement. 

Wyner in {\cite{b1}} proposed the basic model for physical layer security i.e., the wiretap channel, in which the transmitted messages to a legitimate receiver are being overheard by an eavesdropper. Different from the traditional cryptographic techniques {\cite{b2}}, physical layer security utilizes the inherent characteristics of wireless channels to ensure reliable transmission. Recently, researchers devoted to PHY security have shown an increased interest in multiple-input multiple-output
(MIMO) wiretap channels, also referred to as multiple-input multiple-output multiple-eavesdropper (MIMOME) channels {\cite{b6}}, where multiple antennas are deployed at each of the three terminals. The works in {\cite{b4,b5}} investigated the secrecy capacity of MIMOME channels in light of information theory. \cite{b7} extended these work to large-scale systems, which demonstrated the significant improvements of transmission security and reliability in massive MIMOME channel compared to the small-scale one. More specifically, the transmitter can reduce the information disclosure to the eavesdroppers by focusing its main transmit beam to the legitimate receivers\cite{b7}.

Radio-frequency (RF) chain is an expensive component that each antenna should be equipped with, which accounts for high hardware cost in large-scale system. However, antenna selection (AS) technology {\cite{b8}} is regarded as an alternative to alleviate the requirement on the RF transceivers by selecting a subset of antennas to transceive signals. Up to now, the research on AS in MIMOME channels has tended to focus on the closed-form expressions of secrecy capacity under different scenarios but ignore the algorithm design. Most of them merely consider single-antenna selection and the corresponding analytical expression of the secrecy outage probability \cite{b12,b13,b14,b15,b17}. However, a few researches discussed the performance of multiple-antenna selection and very simplistic algorithms were applied to it, such as the norm-based method \cite{b18,b20}. Furthermore, few literatures focused on the algorithm design of multiple-antenna selection in massive MIMOME channels.

This paper concentrates on transmit antenna selection (TAS) algorithm design in massive MIMOME channels. To the best of our knowledge, this is the first time to propose an optimal multiple transmit antenna selection algorithm in massive MIMOME channels, the complexity of which is much lower than that of exhaustive search. For simplicity, assume that the total transmit power is uniformly allocated over the selected antennas and the channel side information of the legitimate receiver (CSIL) is available. An optimal TAS algorithm is proposed to maximize the secrecy capacity in massive MIMOME channels and discussed in two scenarios: 1) For Scenario A: the eavesdropper's channel side information is unavailable at the transmitter (NCSIE), and 2) For Scenario B: the eavesdropper's channel side information (CSIE) is available. In each scenario, simulation results demonstrate that the proposed algorithm obtains an optimal solution at the expense of much lower complexity than exhaustive search. 

The remaining parts of this manuscript is structured as follows: Section \ref{sction2} describes the system model. In Section \ref{sction3}, the optimal TAS algorithm is proposed. The simulation results and corresponding analysis are shown in Section \ref{sction4}. Finally, Section \ref{sction5} concludes the paper.    

$Notations$: Scalars, vectors and matrices are denoted by non-bold, bold lower case, and bold upper letters, respectively. $\mathbb{C}$ stands for the complex numbers. The Hermitian and inverse of matrix $\bf{H}$ is indicated with ${\bf{H}}^{\dagger}$ and ${\bf{H}}^{-1}$, and ${\bf{I}}_N$ is the $N{\times}N$ identity matrix.

\section{System Model} 
\label{sction2}          
In this paper, we consider a massive MIMO wiretap channel. The transmitter is equipped with $N_{\rm{t}}$ antennas, the legitimate receiver is equipped with $N_{\rm{r}}$ antennas and the eavesdropper is equipped with $N_{\rm{e}}$ antennas. The received signal vector at the legitimate receiver reads
\begin{equation}
{{\bf{y}}_{\rm{m}}}=\sqrt{{\rho}_{\rm{m}}}{{\bf{H}}_{\rm{m}}}\bf{x}+{{\bf{w}}_{\rm{m}}},
\end{equation}
where ${\bf{x}}\in{\mathbb{C}}^{N_{\rm{t}}\times1}$ is the transmitted signal with unit power, ${{\rho}_{\rm{m}}}$ is the Signal to Noise Ratio (SNR) at each receive antenna of the legitimate receiver and ${{\bf{w}}_{\rm{m}}}{\sim}{\mathcal{CN}}{({\bf{0}},{{\bf{I}}_{N_{\rm{t}}}})}$ is the additive complex Gaussian noise. Assume that the transmitted symbols from different antennas are independent. Considering independent and identical distributed (i.i.d) Rayleigh flat fading channel, the elements in channel matrix ${\bf{H}_{\rm{m}}}{\in}{{\mathbb{C}}^{{N_{\rm{r}}}{\times}{N_{\rm{t}}}}}$ are i.i.d. complex Gaussian random variables following ${\mathcal{CN}}(0,1)$. Assume that the eavesdropper channel is still suffering i.i.d. Rayleigh flat fading with Gaussian noise. Let $\rho_{\rm{e}}$ denote the SNR at each antenna of the eavesdropper, the received signal in the eavesdropper is given by
\begin{equation}
{{\bf{y}}_{\rm{e}}}=\sqrt{{\rho}_{\rm{e}}}{{\bf{H}}_{\rm{e}}}\bf{x}+{{\bf{w}}_{\rm{e}}}.
\end{equation}
The secrecy capacity between the transmitter and the legitimate receiver is then written as\cite{b5}
\begin{equation}
{{{C}}_{\rm{s}}}=\left[{{C}}_{\rm{m}}-{{C}}_{\rm{e}}\right]^{+},
\end{equation}
where $[x]^+\overset{\triangle}{=}{\rm{max}}\{x,0\}$, ${{C}}_{\rm{e}}$ and ${{C}}_{\rm{m}}$ denote the channel capacity over the eavesdropper channel and the legitimate channel, respectively. Assume that the transmit power is uniformly allocated, ${{C}}_{\rm{e}}$ and ${{C}}_{\rm{m}}$ can be written as \cite{b22}
\begin{subequations}
\begin{align}
{{C}}_{\rm{m}}&=\log_2\det\left({{\bf{I}}_{N_{\rm{t}}}}+{\frac{{{\rho}}_{\rm{m}}}{N_{\rm{t}}}}{{\bf{H}}_{\rm{m}}}{{\bf{H}}{_{\rm{m}}^{\dagger}}}\right)\\
{{C}}_{\rm{e}} &= \log_2\det\left({{\bf{I}}_{N_{\rm{e}}}}+{\frac{{{\rho}}_{\rm{e}}}{N_{\rm{t}}}}{{\bf{H}}_{\rm{e}}}{{\bf{H}}{_{\rm{e}}^{\dagger}}}\right).
\end{align}
\end{subequations}

Then, consider the TAS at the transmitter and suppose $L$ antennas are selected. Actually, selecting a subset of transmit antennas, in other words, is to select the corresponding columns of the channel matrix. Let ${{\tilde{\bf{H}}}_{\rm{m}}}$ and ${{\tilde{\bf{H}}}_{\rm{e}}}$ denote the submatrix after TAS, the secrecy capacity has the following expression:
\begin{equation}
\begin{aligned}
{{{C}}_{\rm{s}}}
&=\left[\log_2\left({\frac{\det\left({{\bf{I}}_{N_{\rm{t}}}}+{{\overline{\rho}}_{\rm{m}}}{{\tilde{\bf{H}}}_{\rm{m}}}{{\tilde{\bf{H}}}{_{\rm{m}}^{\dagger}}}\right)}{\det\left({{\bf{I}}_{N_{\rm{e}}}}+{{\overline{\rho}}_{\rm{e}}}{{\tilde{\bf{H}}}_{\rm{e}}}{{\tilde{\bf{H}}}{_{\rm{e}}^{\dagger}}}\right)}}\right)\right]^{+},
\end{aligned}
\end{equation}
where ${{\overline{\rho}}_{\rm{m}}}=\frac{{{\rho}}_{\rm{m}}}{L}$ and ${{\overline{\rho}}_{\rm{e}}}=\frac{{{\rho}}_{\rm{e}}}{L}$ are defined as the normalized SNR.

\section{TAS Algorithm}
\label{sction3}

In this section, an optimal TAS algorithm with low complexity in massive MIMOME channels is formulated. We assume that full CSIL is available at the transmitter. 

Most of the TAS algorithms used in MIMOME channels are norm-based \cite{b18,b20} i.e., to select $L$ antennas corresponding to the largest $L$ norms of the column vectors in the channel matrix ${\bf{H}}_{\rm{m}}$. The norm-based method is of low complexity but moderately poor performance. Exhaustive search (ES) is definitely an optimal algorithm, but it is prohibitively complex and even impractical for its huge complexity especially under large-scale scenario. Does an optimal algorithm exist for TAS with much lower complexity in contrast to the ES? Branch-and-bound (BAB) method\cite{b23,b24} could answer this question. 

BAB was used for receive antenna selection in massive MIMO system\cite{b24}. In the followings, consider two scenarios stated before depending on whether the CSIE is available or not, and propose corresponding BAB based algorithms for these situations respectively. It's shown that the proposed BAB in Section \ref{scenarioA} is equivalent to that in \cite{b24}, but the algorithm in Section \ref{scenarioB} is totally different. 
\subsection{NCSIE}
\label{scenarioA}
In this case, the transmitter knows nothing about the eavesdropper channel and the transmitted power is uniformly allocated to the selected antennas. Since the CSIE is unavailable at the transmitter, only the legitimate channel is considered in antenna selection. In this setup, the antenna selection for the eavesdropper channel could be treated as a random selection. Let $S$ denote the selected subset of transmit antenna indexes whose cardinality is $|S|=L$. The transmit antenna selection problem could be formulated as
\begin{equation}
{{{S}}^{\rm{opt}}}=\argmax_{{{S}}\in{{\mathcal{M}}}}{\log_2{\det\left({{\bf{I}}_{N_{\rm{r}}}}+{{\overline{\rho}}_{\rm{m}}}{{\tilde{\bf{H}}}_{\rm{m}}}{{\tilde{\bf{H}}}{_{\rm{m}}^{\dagger}}}\right)}}
\end{equation}
where $\mathcal{M}$ denotes the full set of all the candidate column index subsets with size $L$. 

Let ${{\bf{H}}_{{\rm{m}},n}}$ denote the submatrix of legitimate channel after $n$ antennas are selected and ${{{C}}_{{\rm{m}},n}}$ denote the corresponding channel capacity. Assuming that the ${k}$th row ${{\bf{h}}}_{{k}}$ of matrix ${{\bf{H}}_{\rm{m}}}$ is selected in the $\left({n+1}\right)$ step, $n=0,1,\cdots,L-1$, the updated channel submatrix is $\left[{{{\bf{H}}_{{\rm{m}},n}},{{\bf{h}}}_{{k}}}\right]$ the capacity can be derived as
\begin{equation}
\label{eqn_7}
\begin{aligned}
{{{C}}_{{\rm{m}},n+1}}&=\log_2\det\left({{\bf{I}}_{N_{\rm{r}}}}+{{{\overline{\rho}}_{\rm{m}}}}{{\bf{H}}_{{\rm{m}},n+1}}{{\bf{H}}{_{{\rm{m}},n+1}^{\dagger}}}\right)\\
&=\log_2\det\left({{\bf{I}}_{N_{\rm{r}}}}+{{{\overline{\rho}}_{\rm{m}}}}{{\bf{H}}_{{\rm{m}},n}}{{\bf{H}}{_{{\rm{m}},n}}^{\dagger}}+{{{\overline{\rho}}_{\rm{m}}}}{{{\bf{h}}}_{{k}}}{{{\bf{h}}}{_{k}^{\dagger}}}\right)\\
&={{{C}}_{{\rm{m}},n}}+\log_2\det\left({{\bf{I}}_{N_r}}+{{{{\overline{\rho}}_{\rm{m}}}}}{{\bf{T}}_{{\rm{m}},n}}{{{\bf{h}}}_{{k}}}{{{\bf{h}}}{_{k}^{\dagger}}}\right)\\
&\overset{\left(a\right)}{=}{{{C}}_{{\rm{m}},n}}+{\underbrace{\log_2\left(1+{{{{\overline{\rho}}_{\rm{m}}}}}{{{\bf{h}}}{_{k}^{\dagger}}}{{\bf{T}}_{{\rm{m}},n}}{{\bf{h}}_{k}}\right)}_{\Delta_{k,n}}},
\end{aligned}
\end{equation}
where ${{\bf{T}}_{{\rm{m}},n}}=\left({{{\bf{I}}_{N_{\rm{r}}}}+{{{{\overline{\rho}}_{\rm{m}}}}}{{\bf{H}}_{{\rm{m}},n}}{{\bf{H}}{_{{\rm{m}},n}^{\dagger}}}}\right)^{-1}$ and ${{\bf{T}}_{{\rm{m}},0}}={{\bf{I}}_{N_{\rm{r}}}}$. The last equality $\left(a\right)$ holds for the Sylvester’s determinant identity\cite{b25} $\det\left({\bf{I}}+{\bf{A}}{\bf{B}}\right)=\det\left({\bf{I}}+{\bf{B}}{\bf{A}}\right)$. By the Sherman-Morrison formula\cite{b25}, the computation for the matrix inverse $\left({{{\bf{I}}_{N_{\rm{r}}}}+{{{\overline{\rho}}_{\rm{m}}}}{{\bf{H}}_{{\rm{m}},n}}{{\bf{H}}{_{{\rm{m}},n}^{\dagger}}}}\right)^{-1}$ is simplified. Let $K_{n+1}$ denote the antenna index selected in the $\left(n+1\right)$th step, ${{\bf{T}}_{{\rm{m}},n+1}}$ can be
conveniently expressed in the following recursive form
\begin{equation}
\label{eqn_8}
\begin{aligned}
{{\bf{T}}_{{\rm{m}},n+1}}&={\left({{\bf{I}}_{N_{\rm{r}}}}+{{{\overline{\rho}}_{\rm{m}}}}{{\bf{H}}_{{\rm{m}},n}}{{\bf{H}}{_{{\rm{m}},n}^{\dagger}}}+{{{\overline{\rho}}_{\rm{m}}}}{{{\bf{h}}}_{{K_{n+1}}}}{{{\bf{h}}}{_{K_{n+1}}^{\dagger}}}\right)}^{-1}\\
&={{\bf{T}}_{{\rm{m}},n}}-{{{\bf{t}}}_{{n+1}}}{{{\bf{t}}}{_{n+1}^{\dagger}}},
\end{aligned}
\end{equation}
where ${{{\bf{t}}}_{{n+1}}}=\frac{{{\bf{T}}_{{\rm{m}},n}}{{{\bf{h}}}_{{K_{n+1}}}}}{\sqrt{{\left({{{{\overline{\rho}}_{\rm{m}}}}}\right)^{-1}}+{{{{\bf{h}}}{_{K_{n+1}}^{\dagger}}}{{\bf{T}}_{{\rm{m}},n}}{{\bf{h}}_{K_{n+1}}}}}}$. Define $\phi_{{K_{n+1}},n+1}={{{{\bf{h}}}{_{K_{n+1}}^{\dagger}}}{{\bf{T}}_{{\rm{m}},n+1}}{{\bf{h}}_{K_{n+1}}}}$, which can be updated as
\begin{equation}
\begin{aligned}
\phi_{K_{n+1},n+1}&={{{{\bf{h}}}{_{K_{n+1}}^{\dagger}}}{{\bf{T}}_{{\rm{m}},n+1}}{{\bf{h}}_{K_{n+1}}}}\\
&={{{{\bf{h}}}{_{K_{n+1}}^{\dagger}}}{\left({\bf{T}}_{{\rm{m}},n}-{{{\bf{t}}}_{{n+1}}}{{{\bf{t}}}{_{n+1}^{\dagger}}}\right)}{{\bf{h}}_{K_{n+1}}}}\\
&=\phi_{K_{n+1},n}-\left|{\xi}_{K_{n+1},n+1}\right|^2,
\end{aligned}
\end{equation} 
where ${\xi}_{K_{n+1},n+1}={{\bf{h}}{_{K_{n+1}}^{\dagger}}}{{\bf{t}}_{n+1}}$. Therefore, ${\Delta_{k,n}}$ can be written as ${\Delta_{k,n}}=\log_2\left(1+{{{\overline{\rho}}_{\rm{m}}}}{\phi_{k,n}}\right)$.

\begin{figure}[!t] 
\setlength{\abovecaptionskip}{0pt} 
\centering 
\includegraphics[width=0.3\textwidth]{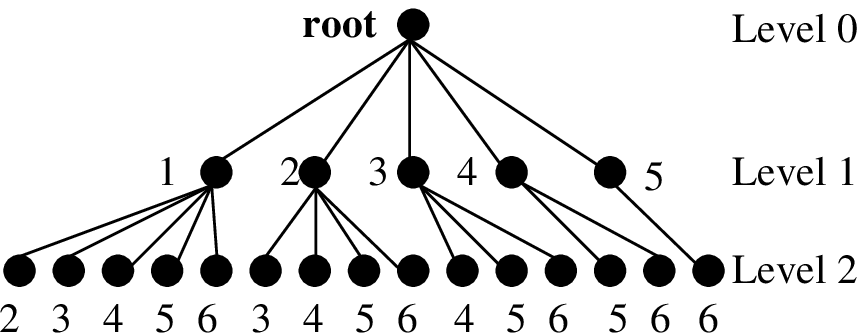} 
\caption{An example search tree for branch-and-bound algorithm when $N_{\rm{t}}=6$ and $L=2$. The number besides each node is the index of this antenna. An entire path from the root to any tip nodes in Level 2 is a antenna selection result.}
\label{tree}
\end{figure}

The BAB search is a classical algorithm in integer programming \cite{b23,b24} which achieves the optimal solution but holds much lower complexity than exhaustive search. To use branch-and-bound search, a search tree as decribed in {\figurename} \ref{tree} is built to implement the whole search. One can see from {\figurename} \ref{tree} that the depth of the search tree is $\left(L+1\right)$. The exhausitive search is to traverse the whole tree. To apply branch-and-bound into TAS, the object function ${{C}}_{{\rm{m}},n}$ is adjusted to $\tilde{{C}}_{{\rm{m}},n}$,
\begin{equation}
\label{equ10}
\tilde{{C}}_{{\rm{m}},n}={{C}}_{{\rm{m}},n}-\sum_{a=0}^{n-1}{Z_a},
\end{equation}
where $Z_a=\log_2\left(1+{{{\overline{\rho}}_m}}\zeta{_a^2}\right)$, and $\zeta{_a^2}=\max_{k\in{\mathcal{I}}_a}{{\bf{h}}{_{k}^{\dagger}}{\bf{h}}_{k}}$, $Z_0=0$. The index set ${\mathcal{I}}_a$ consists of all the candidate antenna set in the $a$th level. In the section that follows, the monotonicity of the new object function $\tilde{{C}}_{{\rm{m}},n}$ is derived in detail. According to the definition of $\tilde{{C}}_{{\rm{m}},n}$, one can see that
\begin{equation}
\tilde{{C}}_{{\rm{m}},n+1}=\tilde{{C}}_{{\rm{m}},n}+\Delta_{{K_{n+1}},n}-{Z_n}.
\end{equation}

Equ. \eqref{eqn_8} shows that ${{\bf{T}}_{{{\rm{m}},n+1}}}={\bf{I}}_{N_{\rm{r}}}-\sum_{a=1}^{n+1}{{\bf{t}}_a}{{\bf{t}}{_a^\dagger}}$ as ${{\bf{T}}_{{\rm{m}},0}}={{\bf{I}}_{N_{\rm{r}}}}$, then
\begin{equation}
\label{equ12}
\begin{aligned}
&{{{{\bf{h}}}{_{K_{n+1}}^{\dagger}}}{{\bf{T}}_{{{\rm{m}},n+1}}}{{\bf{h}}_{K_{n+1}}}}-{{{{\bf{h}}}{_{K_{n+1}}^{\dagger}}}{{\bf{h}}_{K_{n+1}}}}\\
=&-{{{{\bf{h}}}{_{K_{n+1}}^{\dagger}}}\left({\sum_{a=1}^{n+1}{{\bf{t}}_a}{{\bf{t}}{_a^\dagger}}}\right){{\bf{h}}_{K_{n+1}}}}
=-\sum_{a=1}^{n+1}\left|{{{\bf{h}}}{_{K_{n+1}}^{\dagger}}}{{\bf{t}}_a}\right|^2\leq0.
\end{aligned}
\end{equation}
By the definition of ${{Z}}_n$, the relationship ${{{Z}}_n}\geq{\log_2\left(1+{{\overline{\rho}}_m}{\bf{h}}{_{K_{n+1}}^{\dagger}}{{{\bf{h}}}{_{K_{n+1}}}}\right)}$ holds. As a result, 
\begin{equation}
\label{equ13}
\begin{aligned}
\Delta_{{K_{n+1}},n}&=\log_2\left(1+{{\overline{\rho}}_m}{{{{\bf{h}}}{_{K_{n+1}}^{\dagger}}}{{\bf{T}}_{{{\rm{m}},n+1}}}{{\bf{h}}_{K_{n+1}}}}\right)\\
&\leq{\log_2\left(1+{{\overline{\rho}}_m}{{{{\bf{h}}}{_{K_{n+1}}^{\dagger}}}{{\bf{h}}_{K_{n+1}}}}\right)}\leq{{{{Z}}_n}}.
\end{aligned}
\end{equation}
By Equ. \eqref{equ10} and Equ. \eqref{equ13}, $\tilde{{C}}_{{\rm{m}},n+1}\leq\tilde{{C}}_{{\rm{m}},n}$ holds, which indicates that $\tilde{{C}}_{{\rm{m}},n}$ is monotonically decreasing with the increase of the number of the selected antennas. 

As ${{{Z}}_n}$ is the maximal value in each level, they are constants once the search tree is fixed. Therefore, maximizing ${{C}}_{{\rm{m}},n}$ is equivalent to maximizing $\tilde{{C}}_{{\rm{m}},n}$ by Equ. \eqref{equ10}. Branch-and-bound search is suitable to find maximum with a monotonically-decreasing object function\cite{b23}. Suppose that the depth-first and best-first strategy is used during tree search. Since the object function $\tilde{{C}}_{{\rm{m}},n}$ along a path is decreasing, the object function value of a complete path from the root node to the tip node could serve as a lower bound for other nodes. 
For example, when a node '$A$' of a path in the $q$th level is visited, all the child nodes produced by '$A$' can be discard if the real-time object function value $\tilde{{C}}_{{\rm{m}},q}$ is smaller than the lower bound. 
When we arrive at another tip node, we need to update the lower bound as the object function value of this new complete path if the object function value is larger than the lower bound. This procedure will not stop until the whole tree is traversed. Let minus infinity be the initial global lower bound. The tighter the initial bound is, the lower complexity the BAB algorithm would possess. The branch-and-bound algorithm is summarized in Alg. \ref{alg1} attached with the complexity analysis, where ${\mathcal{I}}_{n,{k}}$ is the sub-node index set in the $n$th level of the $k$th node.

\begin{algorithm}[!t]
\caption{BAB search for TAS with NCSIE \cite{b24}}
\label{alg1}
\begin{algorithmic}[1]
\State ${\bf{T}}_{{\rm{m}}}={\bf{I}}_{{N_{\rm{r}}}}$, $B=-\infty$, $\tilde{{C}}_{{\rm{m}}}=0$, $n=1$, $K=0$, ${\bf{s=0}}_{{L}}$, ${{\mathcal{K}}}=\{1,2,\cdots,N_{\rm{t}}\}$, ${{\mathcal{L}}}=\{1,2,\cdots,L\}$ 
\State $\phi_{k}={\bf{h}}{_{k}^{\dagger}}{\bf{h}}_{k},~\forall{k}\in{{\mathcal{K}}}$~~~~~$\rhd~{\mathcal{O}}\left(N_{\rm{t}}N_{\rm{r}}\right)$
\State $\zeta_a=\max_{k\in{\mathcal{I}}_a}\phi_{k}, Z_a=\log_2\left(1+{\overline{{\rho}}_{\rm{m}}}{\zeta_a}\right), \forall{a}\in{\mathcal{L}}$
\State $\Delta_{k}=\log_2\left(1+{\overline{{\rho}}_{\rm{m}}}\phi_{k}\right), \forall{k}\in{{\mathcal{K}}}$
\If{$n=L$}
\State ${c_k}\coloneqq\tilde{{C}}_{{\rm{m}}}+\Delta_{k}-Z_{L}, \forall{k}\in{\mathcal{I}}_{L,{K}}$
\If{$\max_{{m}\in{\mathcal{I}}_{L,{K}}}{c_m}<B$}
\State $\left[{\bf{s}}\right]_L=\arg\max_{{m}\in{\mathcal{I}}_{L,{K}}}{c_m}$
\State $B\coloneqq\max_{{m}\in{\mathcal{I}}_{L,{K}}}{c_m}$, and $\hat{{\bf{s}}}\coloneqq{{{\bf{s}}}}$
\EndIf
\Else
\State ${c_k}\coloneqq\tilde{{C}}_{{\rm{m}}}+\Delta_{k}-Z_{n}, \forall{j}\in{\mathcal{I}}_{n,{K}}$
\State sort ${c_k}, \forall{k}\in{\mathcal{I}}_{n,{K}}$ in a descend order to get an ordered 

index vector ${\bf{k}}$
\State ${{\bf{T}}_{{\rm{m}},tmp}}\coloneqq{{\bf{T}}_{{\rm{m}}}}$, $\phi_{tmp,k}\coloneqq\phi_{k}$, $\forall{k}\in{\mathcal{K}}$
\For{$i=1:\left|{\mathcal{I}}_{n,{K}}\right|$}
\State ${K}=\left[{\bf{k}}\right]_i$
\If{$c_{K}>B$}
\State $\mathcal{Q}\coloneqq\{K+1,K+2,\cdots,N_{\rm{t}}\}$
\State $\left[{\bf{s}}\right]_{n}={K}$
\State ${{\bf{t}}_{\rm{m}}}\coloneqq\frac{{{\bf{T}}_{{\rm{m}}}}{{{\bf{h}}}_{{K}}}}{\sqrt{{\left({\overline{{\rho}}_{\rm{m}}}\right)^{-1}}+{\phi_{tmp,{K}}}}}$~~~~~$\rhd~{\mathcal{O}}\left(N_{\rm{t}}^2N_{\text{nodes}}\right)$
\State ${{\bf{T}}_{{\rm{m}}}}\coloneqq{{\bf{T}}_{{\rm{m}},tmp}}-{{\bf{t}}_{\rm{m}}}{{\bf{t}}{_{\rm{m}}^\dagger}}$
\State $\tilde{{C}}_{{\rm{m}}}\coloneqq{c_{K}}$
\State ${\xi}_{a}\coloneqq{{\bf{h}}{_{a}^{\dagger}}}{{\bf{t}}_{\rm{m}}},\forall{a\in{{\mathcal{Q}}}}$~~~~~$\rhd~{\mathcal{O}}\left(N_{\rm{t}}N_{\rm{r}}N_{\text{nodes}}\right)$
\State ${\phi_{a}}\coloneqq{\phi_{tmp,a}}-\left|{{\xi}_{a}}\right|^2,\forall{a\in{{\mathcal{Q}}}}$~~$\rhd~{\mathcal{O}}\left(N_{\rm{t}}N_{\text{nodes}}\right)$
\State ${\Delta_a}\coloneqq\log_2\left(1+{\overline{{\rho}}_{\rm{m}}}{\phi_{a}}\right),\forall{a\in{{\mathcal{Q}}}}$

{\qquad}{\qquad}{\qquad}{\qquad}{\qquad}{\qquad}{\qquad}$\rhd~{\mathcal{O}}\left(N_{\rm{t}}N_{\rm{r}}N_{\text{nodes}}\right)$
\State $n\coloneqq{n+1}$, jump to line 5
\Else
\State break the loop
\EndIf
\EndFor
\EndIf\\
\Return the final set $\hat{{\bf{s}}}$
\end{algorithmic}
\end{algorithm}

$\tilde{{C}}_{{\rm{m}},n}$ is regarded as the objective function instead of ${{C}}_{{\rm{m}},n}$, because the former is monotonically decreasing. During the procedure of tree search, many nodes could be pruned. Suppose ${{C}}_{{\rm{m}},n}$ was utilized, the whole procedure would degrade into an exhaustive search with huge computational complexity. If the tree search stops at the first level, it would degrade into the norm-based method. During the procedure of branch-and-bound algorithm, the total number of visited node is uncertain, the computational complexity could be calculated by ${\mathcal{O}}\left(N_{\text{node}}N_{\rm{t}}N_{\rm{r}}\right)$ where $N_{\text{nodes}}$ denotes the visited nodes' total number. Many branches are pruned during the algorithm, thuis it could achieve the optimal solution with much lower computation cost than exhaustive search.

\subsection{CSIE}
\label{scenarioB}
The previous section has discussed the scenario without CSIE. Then, consider the situation when the transmitter have both full CSIL and full CSIE. The following part moves on to describe in greater detail that the BAB still works well when CSIE is available.

Since the transmitter knows the channel matrix of the eavesdropper, the antenna selection for the eavesdropper channel can not be treated as random selection any more and the channel side information must be taken into consideration. Let ${{\bf{H}}_{{\rm{m}},n}}$ and ${{\bf{H}}_{{\rm{e}},n}}$ denote the submatrix of legitimate channel and eavesdropper channel after $n$ antennas are selected and ${{{C}}_{{\rm{s}},n}}$ denote the corresponding secrecy capacity. Assuming that the ${k}$th column ${{\bf{h}}}_{{\rm{m}},k}$ of matrix ${{\bf{H}}_{\rm{m}}}$ is selected at the $\left({n+1}\right)$ step, the channel matrix is denoted by $\left[{{{\bf{H}}_{{\rm{m}},n}},{{\bf{h}}}_{{k}}}\right]$. For the eavesdropper, the ${k}$th column ${{\bf{h}}}_{{\rm{e}},k}$ of matrix ${{\bf{H}}_{\rm{e}}}$ is selected. 
Following the similar steps in Equ. \eqref{eqn_7} yields another recursive formulation
\begin{equation}
{{{C}}_{{\rm{s}},n+1}}={{{C}}_{{\rm{s}},n}}+{\underbrace{\log_2\left(\frac{1+{\overline{{\rho}}_{\rm{m}}}{\bf{h}}{_{{\rm{m}},k}^{\dagger}}{{\bf{T}}_{{\rm{m}},n}}{{\bf{h}}_{{\rm{m}},k}}}{1+{\overline{{\rho}}_{\rm{e}}}{{{\bf{h}}}{_{{\rm{e}},k}^{\dagger}}}{{\bf{T}}_{{\rm{e}},n}}{{\bf{h}}_{{\rm{e}},k}}}\right)}_{\Delta_{k,n}}},
\end{equation} 
where ${{\bf{T}}_{{\rm{m}},n}}={\left({{\bf{I}}_{N_{\rm{r}}}}+{\overline{{\rho}}_{\rm{m}}}{{\bf{H}}_{{\rm{m}},n}}{{\bf{H}}{_{{\rm{m}},n}^{\dagger}}}\right)}^{-1}$ and ${{\bf{T}}_{{\rm{e}},n}}={\left({{\bf{I}}_{N_{\rm{e}}}}+{\overline{{\rho}}_{\rm{e}}}{{\bf{H}}_{{\rm{e}},n}}{{\bf{H}}{_{{\rm{e}},n}^{\dagger}}}\right)}^{-1}$. By the Sherman-Morrison formula, these expressions are simplified as follows:
\begin{subequations}
\begin{align}
{{\bf{T}}_{{\rm{m}},n+1}}
&={{\bf{T}}_{{\rm{m}},n}}-{{{\bf{t}}}_{{{\rm{m}},n+1}}}{{{\bf{t}}}{_{{\rm{m}},n+1}^{\dagger}}},\\
{{\bf{T}}_{{\rm{e}},n+1}}
&={{\bf{T}}_{{\rm{e}},n}}-{{{\bf{t}}}_{{{\rm{e}},n+1}}}{{{\bf{t}}}{_{{\rm{e}},n+1}^{\dagger}}},
\end{align}
\end{subequations}
in which ${{{\bf{t}}}_{{\rm{m}},n+1}}$ and ${{{\bf{t}}}_{{\rm{e}},n+1}}$ hold the similar expressions as $t_{n+1}$ in Section \ref{scenarioA}.  

Define 
\begin{subequations}
\begin{align}
\phi_{{\rm{m}},k,n+1}&=\phi_{{\rm{m}},k,n}-\left|{\xi}_{{\rm{m}},k,n+1}\right|^2,\\
\phi_{{{\rm{e}},k,n+1}}&=\phi_{{\rm{e}},k,n}-\left|{\xi}_{{\rm{e}},j,n+1}\right|^2,
\end{align}
\end{subequations}
where ${\xi}_{{\rm{m}},k,n+1}={{\bf{h}}{_{{\rm{m}},k}^{\dagger}}}{{\bf{t}}_{{\rm{m}},n+1}}$ and ${\xi}_{{\rm{e}},k,n+1}={{\bf{h}}{_{{\rm{e}},k}^{\dagger}}}{{\bf{t}}_{{\rm{e}},n+1}}$. 
Then, ${\Delta_{k,n}}=\log_2\left(1+{\overline{{\rho}}_{\rm{m}}}{\phi_{{\rm{m}},k,n}}\right)-\log_2\left(1+{\overline{{\rho}}_{\rm{e}}}{\phi_{{\rm{e}},k,n}}\right)$. $C_{{\rm{s}},L}$ should be adjusted to a monotonically-decreasing function, that is
\begin{equation}
\tilde{{C}}_{{\rm{s}},n}={{C}}_{{\rm{s}},n}-\sum_{a=0}^{n-1}{Z_a},
\end{equation}
where $Z_a=\log_2\left(1+{\overline{{\rho}}_{\rm{m}}}\zeta{_a^2}\right)-\log_2\left(1+{\overline{{\rho}}_{\rm{e}}}\eta_a\right)$, in which $\zeta{_a^2}=\max_{k\in{\mathcal{I}}_a}{{\bf{h}}{_{{\rm{m}},k}^{\dagger}}{\bf{h}}_{{\rm{m}},k}}$ and $\eta_a$ is defined as $\eta_a=\min_{k\in{\mathcal{I}}_a}{{\bf{h}}{_{{\rm{e}},k}^{\dagger}}{{\left({{\bf{I}}_{N_{\rm{e}}}}+{\overline{{\rho}}_{\rm{e}}}{{\bf{H}}_{\rm{e}}}{{\bf{H}}{_{\rm{e}}^{\dagger}}}\right)}^{-1}}{\bf{h}}_{{\rm{e}},k}}$. The index set ${\mathcal{I}}_a$ consists of all the candidate antenna set in the $a$th level, and $Z_0=0$. Suppose that the ${K_{n+1}}$th antenna has been selected in the $\left(n+1\right)$th step, a recursive formula is derived, namely
\begin{equation}
\label{equ18}
\tilde{{C}}_{{\rm{s}},n+1}=\tilde{{C}}_{{\rm{s}},n}+\Delta_{{K_{n+1}},n}-{Z_n}.
\end{equation}
Since ${{\bf{T}}_{{{\rm{e}},n+1}}}={\bf{I}}_{N_{\rm{e}}}-\sum_{a=1}^{n+1}{{\bf{t}}_{{\rm{e}},a}}{{\bf{t}}{_{{\rm{e}},a}^\dagger}}$, once all the $N_{\rm{t}}$ antennas are selected, ${{\bf{T}}_{{{\rm{e}},N_{\rm{t}}}}}$ is a constant by its definition i.e., ${{\bf{T}}_{{{\rm{e}},N_{\rm{t}}}}}={\bf{I}}_{N_{\rm{e}}}-\sum_{a=1}^{N_{\rm{e}}}{{\bf{t}}_{{\rm{e}},a}}{{\bf{t}}{_{{\rm{e}},a}^\dagger}}={{\left({{\bf{I}}_{N_{\rm{e}}}}+{\overline{{\rho}}_{\rm{e}}}{{\bf{H}}_{\rm{e}}}{{\bf{H}}{_{\rm{e}}^{\dagger}}}\right)}^{-1}}$, which shows that ${\bf{T}}_{{{\rm{e}},N_{\rm{t}}}}$ is fixed in any selection order. Therefore,
\begin{equation}
\label{eq19}
\begin{aligned}
&{{{{\bf{h}}}{_{{\rm{e}},K_{n+1}}^{\dagger}}}{{\bf{T}}_{{{\rm{e}},n+1}}}{{\bf{h}}_{{\rm{e}},K_{n+1}}}}-{{{{\bf{h}}}{_{{\rm{e}},K_{n+1}}^{\dagger}}}{{\bf{T}}_{{{\rm{e}},N_{\rm{e}}}}}{{\bf{h}}_{{\rm{e}},K_{n+1}}}}\\
\overset{(a)}{=}&{{{{\bf{h}}}{_{K_{n+1}}^{\dagger}}}\left({\sum_{a=n+2}^{N_{\rm{e}}}{{\bf{t}}_a}{{\bf{t}}{_a^\dagger}}}\right){{\bf{h}}_{K_{n+1}}}}
=\sum_{a=n+2}^{N_{\rm{e}}}\left|{{{\bf{h}}}{_{K_{n+1}}^{\dagger}}}{{\bf{t}}_a}\right|^2\geq0.
\end{aligned}
\end{equation}
The step ($a$) in Equ. \eqref{eq19} holds for that one could treat ${{\bf{T}}_{{{\rm{e}},N_{\rm{e}}}}}$ as the result of any selection order which includes the first $\left(n+1\right)$ antennas index that results in ${{\bf{T}}_{{{\rm{e}},n+1}}}$.

According to the definition of $\eta_a$, $\eta_a=\min_{k\in{\mathcal{I}}_a}{{\bf{h}}{_{{\rm{e}},k}^{\dagger}}{{{\bf{T}}_{{{\rm{e}},N_{\rm{e}}}}}}{\bf{h}}_{{\rm{e}},k}}$, thus $\eta_a\leq{{{{{\bf{h}}}{_{{\rm{e}},K_{n+1}}^{\dagger}}}{{\bf{T}}_{{{\rm{e}},K_{n+1}}}}{{\bf{h}}_{{\rm{e}},K_{n+1}}}}}$. In addition, it has been proved in Equ. \eqref{equ12} and \eqref{equ13} that ${{{{{\bf{h}}}{_{{\rm{m}},K_{n+1}}^{\dagger}}}{{\bf{T}}_{{{\rm{m}},N_{\rm{m}}}}}{{\bf{h}}_{{\rm{m}},K_{n+1}}}}}\leq{\zeta{_a^2}}$. Consequently,
\begin{equation}
\begin{aligned}
\Delta_{{K_{n+1}},n}&=\log_2\left(\frac{1+{{\overline{\rho}}_m}{{{{\bf{h}}}{_{{{\rm{m}}},K_{n+1}}^{\dagger}}}{{\bf{T}}_{{{\rm{m}},n+1}}}{{\bf{h}}_{{{\rm{m}}},K_{n+1}}}}}{1+{{\overline{\rho}}_e}{{{{\bf{h}}}{_{{{\rm{e}}},K_{n+1}}^{\dagger}}}{{\bf{T}}_{{{\rm{e}},n+1}}}{{\bf{h}}_{{{\rm{e}}},K_{n+1}}}}}\right)\\
&\leq{{\log_2\left(1+{{\overline{\rho}}_m}\zeta{_a^2}\right)}-{\log_2\left(1+{\overline{{\rho}}_{\rm{e}}}\eta_a\right)}}
={{{{Z}}_n}}.
\end{aligned}
\end{equation}
Thus $\tilde{{C}}_{{\rm{s}},n+1}\leq\tilde{{C}}_{{\rm{s}},n}$ holds by Equ. \eqref{equ18}, which indicates that $\tilde{{C}}_{{\rm{s}},n+1}$ is monotonically decreasing.  
It's clear that the branch-and-bound for the situation with full CSIE is different from the one with NCSIE. In Section \ref{scenarioA}, the monotonically-increasing function $C_{{\rm{m}},n}$ is adjusted to a monotonically-decreasing function $\tilde{{C}}_{{\rm{m}},n}$. Nevertheless, the original function $C_{{\rm{s}},n}$ isn't monotonic, which makes it even harder to do the construction. The branch-and-bound with full CSIE is summarized in Alg. \ref{alg2}, where ${\mathcal{I}}_{n,{k}}$ is the subnode index set of the $j$th node in the $n$th level. Similar as Alg. \ref{alg1}, the complexity of Alg. \ref{alg2} is ${\mathcal{O}}\left(N_{\text{nodes}}N_{\rm{t}}\max\left(N_{\rm{r}},N_{\rm{e}}\right)\right)$ where $N_{\text{nodes}}$ denotes the total number of the visited nodes,
\begin{algorithm}[!t]
\caption{BAB search for TAS with CSIE.}
\label{alg2}
\begin{algorithmic}[1]
\State ${\bf{T}}_{\rm{m}}={\bf{I}}_{N_{\rm{r}}}$, ${\bf{T}}_{\rm{e}}={\bf{I}}_{N_{\rm{e}}}$, $B=-\infty$, $\tilde{{C}}_{\rm{s}}=0$, $n=1$, $K_0=0$, ${\bf{s=0}}_{{L}}$, ${{\mathcal{K}}}=\{1,2,\cdots,N_{\rm{t}}\}$, ${{\mathcal{L}}}=\{1,2,\cdots,L\}$
\State $\phi_{{\rm{m}},k}={\bf{h}}{_{{\rm{m}},k}^{\dagger}}{\bf{h}}_{{\rm{m}},k}, \phi_{{\rm{e}},k}={\bf{h}}{_{{\rm{e}},k}^{\dagger}}{\bf{h}}_{{\rm{e}},k},\forall{k}\in{{\mathcal{K}}}$
\State $\zeta_a=\max_{k\in{\mathcal{I}}_a}\phi_{{\rm{m}},k}, \forall{a}\in{\mathcal{L}}$\\ 
$\eta_a=\min_{k\in{\mathcal{I}}_a}{{\bf{h}}{_{{\rm{e}},k}^{\dagger}}{{\left({{\bf{I}}_{N_{\rm{e}}}}+{{\overline{\rho}}_{\rm{e}}}{{\bf{H}}_{\rm{e}}}{{\bf{H}}{_{\rm{e}}^{\dagger}}}\right)}^{-1}}{\bf{h}}_{{\rm{e}},k}}, \forall{a}\in{\mathcal{L}}$
\State $Z_a=\log_2\left(1+{\overline{{\rho}}_{\rm{m}}}\zeta{_a^2}\right)-\log_2\left(1+{\overline{{\rho}}_{\rm{e}}}\eta_a\right), \forall{a}\in{\mathcal{L}}$
\State $\Delta_{k}=\log_2\left(1+{\overline{{\rho}}_{\rm{m}}}\phi_{{\rm{m}},k}\right)-\log_2\left(1+{\overline{{\rho}}_{\rm{e}}}\phi_{{\rm{e}},k}\right), \forall{k}\in{{\mathcal{K}}}$, 
\If{$n=L$}
\State ${c_k}\coloneqq\tilde{{C}}_{{\rm{s}}}+\Delta_{k}-Z_{L}, \forall{k}\in{\mathcal{I}}_{L,{K}}$
\If{$\max_{{m}\in{\mathcal{I}}_{L,{K}}}{c_m}<B$}
\State $\left[{\bf{s}}\right]_L=\arg\max_{{m}\in{\mathcal{I}}_{L,{K}}}{c_m}$
\State $B\coloneqq\max_{{m}\in{\mathcal{I}}_{L,{K}}}{c_m}$, and $\hat{{\bf{s}}}\coloneqq{{{\bf{s}}}}$
\EndIf
\Else
\State ${c_k}\coloneqq\tilde{{C}}_{{\rm{s}}}+\Delta_{k}-Z_{n}, \forall{k}\in{\mathcal{I}}_{n,{K}}$
\State sort ${c_k}, \forall{k}\in{\mathcal{I}}_{n,{K}}$ in a descend order to get an ordered 

index vector $\bf{k}$
\State 
${{\bf{T}}_{{\rm{m}},tmp}}{\coloneqq}{{\bf{T}}_{{\rm{m}}}}$, $\phi_{{\rm{m}},tmp,k}{\coloneqq}\phi_{{\rm{m}},k}$, $\forall{k}\in{\mathcal{K}}$

${{\bf{T}}_{{\rm{e}},tmp}}{\coloneqq}{{\bf{T}}_{{\rm{e}}}}$, $\phi_{{\rm{e}},tmp,k}{\coloneqq}\phi_{{\rm{e}},k}$, $\forall{k}\in{\mathcal{K}}$
\For{$i=1:\left|{\mathcal{I}}_{n,{K}}\right|$}
\State ${K}=\left[{\bf{k}}\right]_i$
\If{$c_{K}>B$}
\State $\mathcal{Q}\coloneqq\{K+1,K+2,\cdots,N_{\rm{t}}\}$
\State $\left[{\bf{s}}\right]_{n+1}={K}$
\State ${\bf{t}}_{\rm{m}}{\coloneqq}\frac{{{\bf{T}}_{\rm{m}}}{{{\bf{h}}}_{{\rm{m}},{K}}}}{\sqrt{{\left({{\overline{\rho}}_{\rm{m}}}\right)^{-1}}+{\phi_{{\rm{m}},tmp,{K}}}}}$

~~~~~~~${{\bf{T}}_{{\rm{m}}}}{\coloneqq}{{\bf{T}}_{{\rm{m}},tmp}}-{{\bf{t}}_{{\rm{m}}}}{{\bf{t}}{_{{\rm{m}}}^\dagger}}$

~~~~~~~${\bf{t}}_{\rm{e}}{\coloneqq}\frac{{{\bf{T}}_{\rm{e}}}{{{\bf{h}}}_{{\rm{e}},{K}}}}{\sqrt{{\left({{\overline{\rho}}_{\rm{e}}}\right)^{-1}}+{\phi_{{\rm{e}},tmp,{K}}}}}$

~~~~~~~${{\bf{T}}_{{\rm{e}}}}{\coloneqq}{{\bf{T}}_{{\rm{e}},tmp}}-{{\bf{t}}_{{\rm{e}}}}{{\bf{t}}{_{{\rm{e}}}^\dagger}}$
\State $\tilde{{C}}_{{\rm{s}},n}{\coloneqq}c_{K}$
\State ${\xi}_{{\rm{m}},a}{\coloneqq}{{\bf{h}}{_{{\rm{m}},a}^{\dagger}}}{{\bf{t}}_{{\rm{m}}}}, {\xi}_{{\rm{e}},a}{\coloneqq}{{\bf{h}}{_{{\rm{e}},a}^{\dagger}}}{{\bf{t}}_{{\rm{e}}}}, \forall{a\in{{\mathcal{Q}}}}$
\State ${\phi_{{\rm{m}},a}}{\coloneqq}{\phi_{{\rm{m}},tmp,a}}-\left|{{\xi}_{{\rm{m}},a}}\right|^2, \forall{a\in{{\mathcal{Q}}}}$ 

~~~~~~~${\phi_{{\rm{e}},a}}{\coloneqq}{\phi_{{\rm{e}},tmp,a}}-\left|{{\xi}_{{\rm{e}},a}}\right|^2, \forall{a\in{{\mathcal{Q}}}}$
\State ${\Delta_a}{\coloneqq}\log_2\left(\frac{1+{\overline{\rho}_{\rm{m}}}{\phi_{{\rm{m}},a}}}{1+{\overline{\rho}_{\rm{e}}}{\phi_{{\rm{e}},a}}}\right), \forall{a\in{{\mathcal{Q}}}}$
\State $n\coloneqq{n+1}$, jump to line 7
\Else
\State break the loop
\EndIf
\EndFor
\EndIf\\
\Return the final set $\hat{{\bf{s}}}$
\end{algorithmic}
\end{algorithm}

\section{Simulation Results}
\label{sction4}
This part gives the simulation results followed by the computation complexity analysis of all the proposed algorithms.

\begin{figure}[!t] 
\setlength{\abovecaptionskip}{-5pt} 
\centering 
\includegraphics[width=0.40\textwidth]{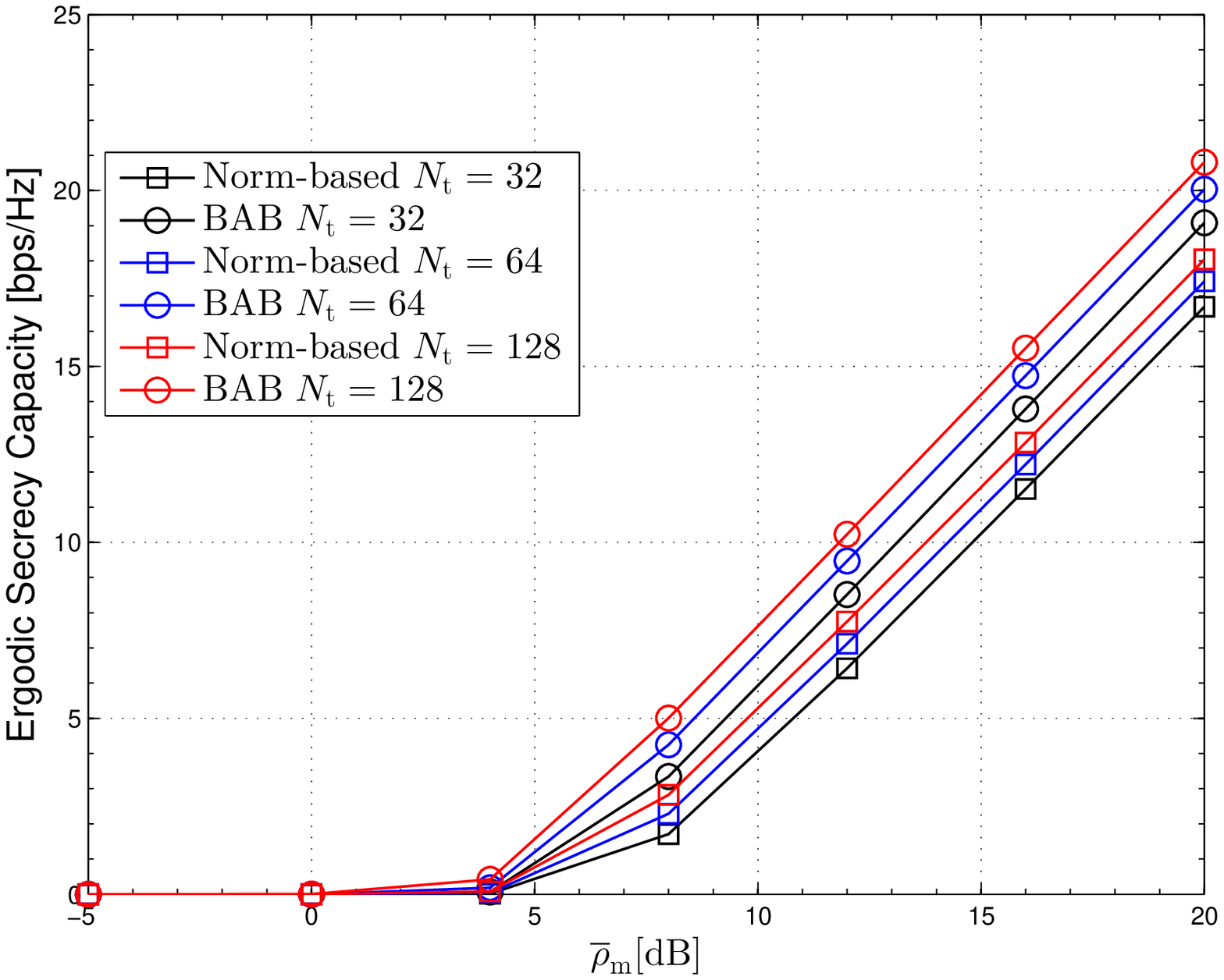} 
\caption{Ergodic secrecy capacity verus ${\overline{\rho}}_{\rm{m}}$ using BAB-based and norm-based method with NCSIE, $N_r=4, N_e=8, L=4$ and ${\overline{\rho}}_{\rm{e}}=5\text{dB}$.}
\label{figure1}
\end{figure}

\begin{figure}[!t] 
\setlength{\abovecaptionskip}{-5pt} 
\centering 
\includegraphics[width=0.40\textwidth]{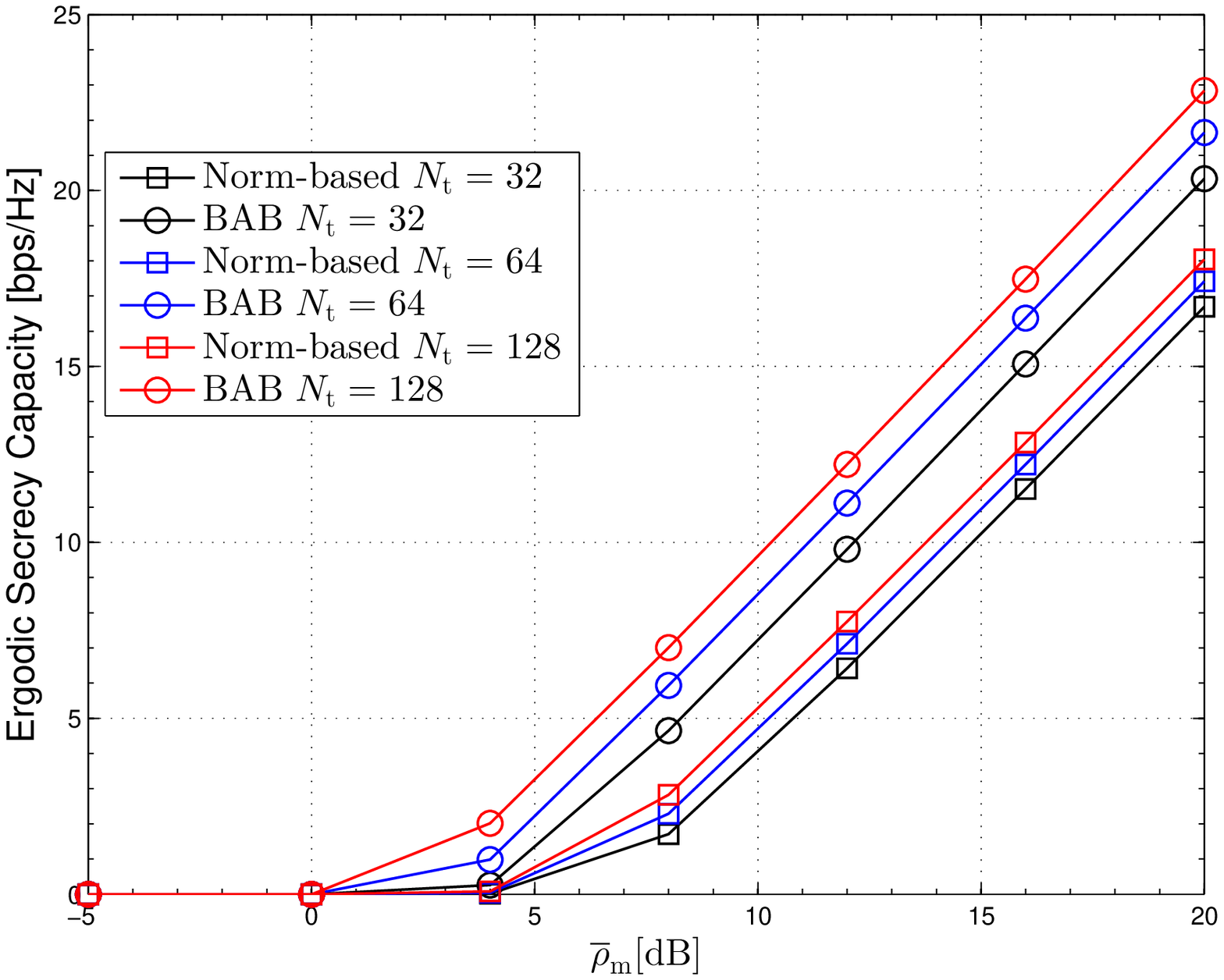} 
\caption{Ergodic secrecy capacity verus ${\overline{\rho}}_{\rm{m}}$ using BAB-based and norm-based method with CSIE, $N_r=4, N_e=8, L=4$ and ${\overline{\rho}}_{\rm{e}}=5\text{dB}$.}
\label{figure2}
\end{figure}

\begin{figure}[!t] 
\setlength{\abovecaptionskip}{-5pt} 
\centering 
\includegraphics[width=0.40\textwidth]{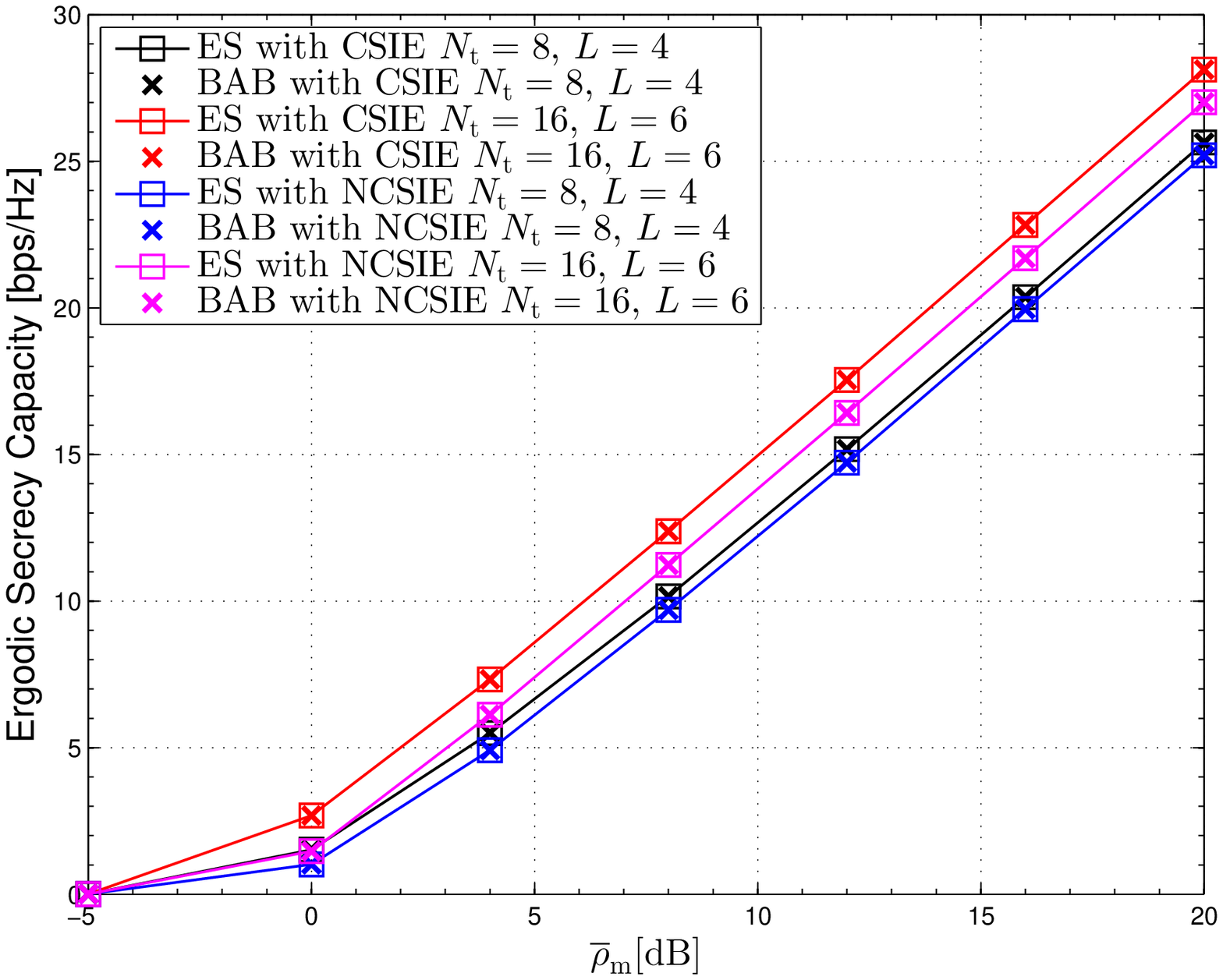} 
\caption{Ergodic secrecy capacity verus ${\overline{\rho}}_{\rm{m}}$ using BAB and ES, $N_r=4, N_e=4$ and ${\overline{\rho}}_{\rm{e}}=1\text{dB}$.}
\label{figure3}

\end{figure}

{\figurename} \ref{figure1} presents the ergodic secrecy capacity versus $\overline{\rho}_{\rm{m}}$ when CSIE is unavailable at the transmitter. 
As is shown in {\figurename} \ref{figure1}, BAB based search has superior performance in all conditions compared with the norm-based search. Furthermore, {\figurename} {\ref{figure2}} shows the ergodic secrecy capacity for BAB method and norm-based method when CSIE is available. Also, it can be seen from the figure that BAB still outperforms the norm based method. Finally, to verify that the branch-and-bound search has the same optimal performance as exhaustive search, {\figurename} \ref{figure3} compares the ergodic secrecy capacity of BAB and ES in both NCSIE and CSIE cases. It is apparent from {\figurename} \ref{figure3} that the BAB search can find the optimal antenna index subset to maximize the secrecy capacity.

\setlength{\floatsep}{5pt}
\begin{figure}[!t] 
\setlength{\abovecaptionskip}{-5pt} 
\centering 
\includegraphics[width=0.40\textwidth]{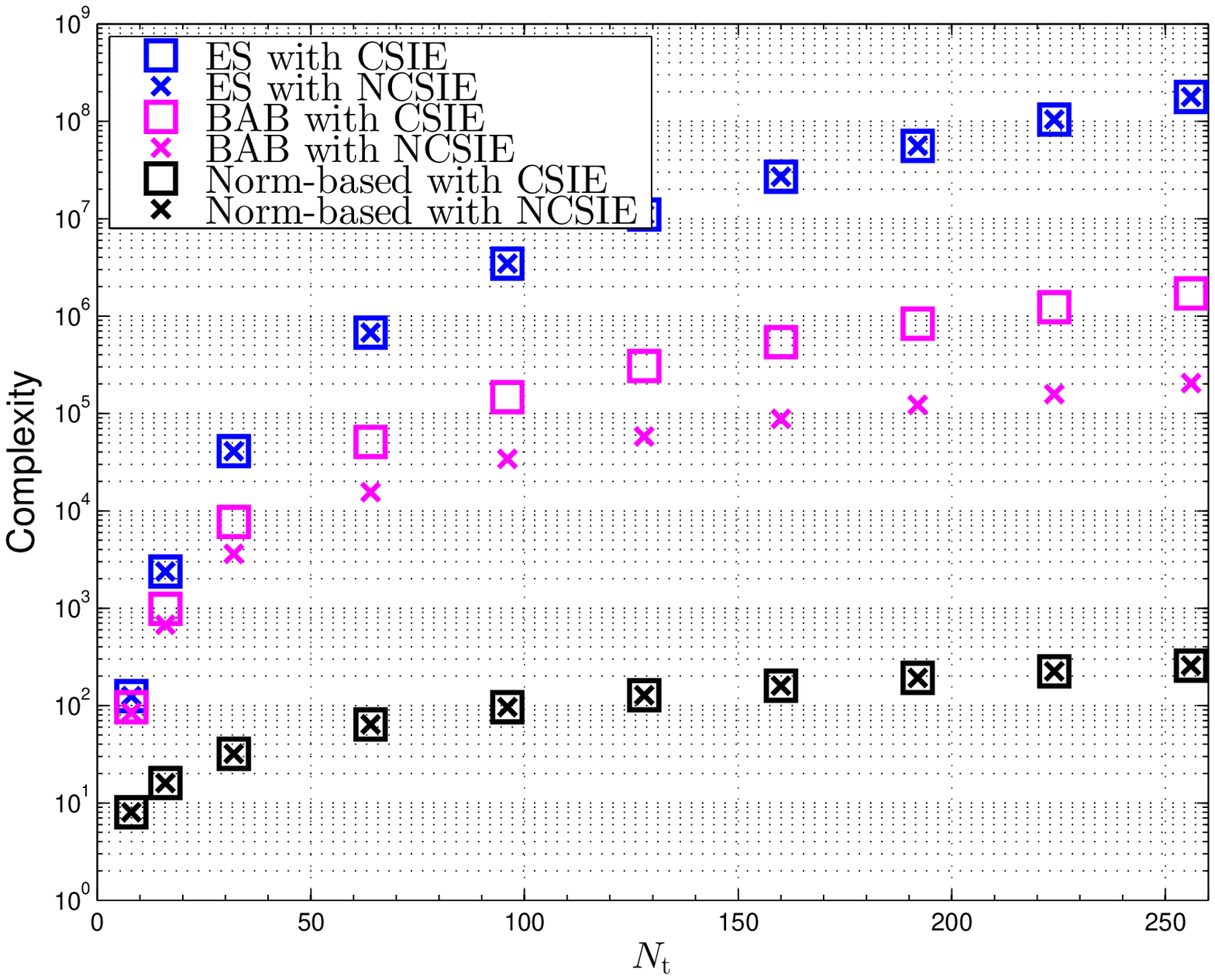} 
\caption{Complexity of BAB \& Norm-based \& ES verus $N_t$, $N_r=4, N_e=4, L=4, {\overline{\rho}}_{\rm{m}}=9\text{dB}$ and ${\overline{\rho}}_{\rm{e}}=1\text{dB}$.}
\label{figure4}
\end{figure}

\begin{figure}[!t] 
\setlength{\abovecaptionskip}{-5pt}
\centering 
\includegraphics[width=0.40\textwidth]{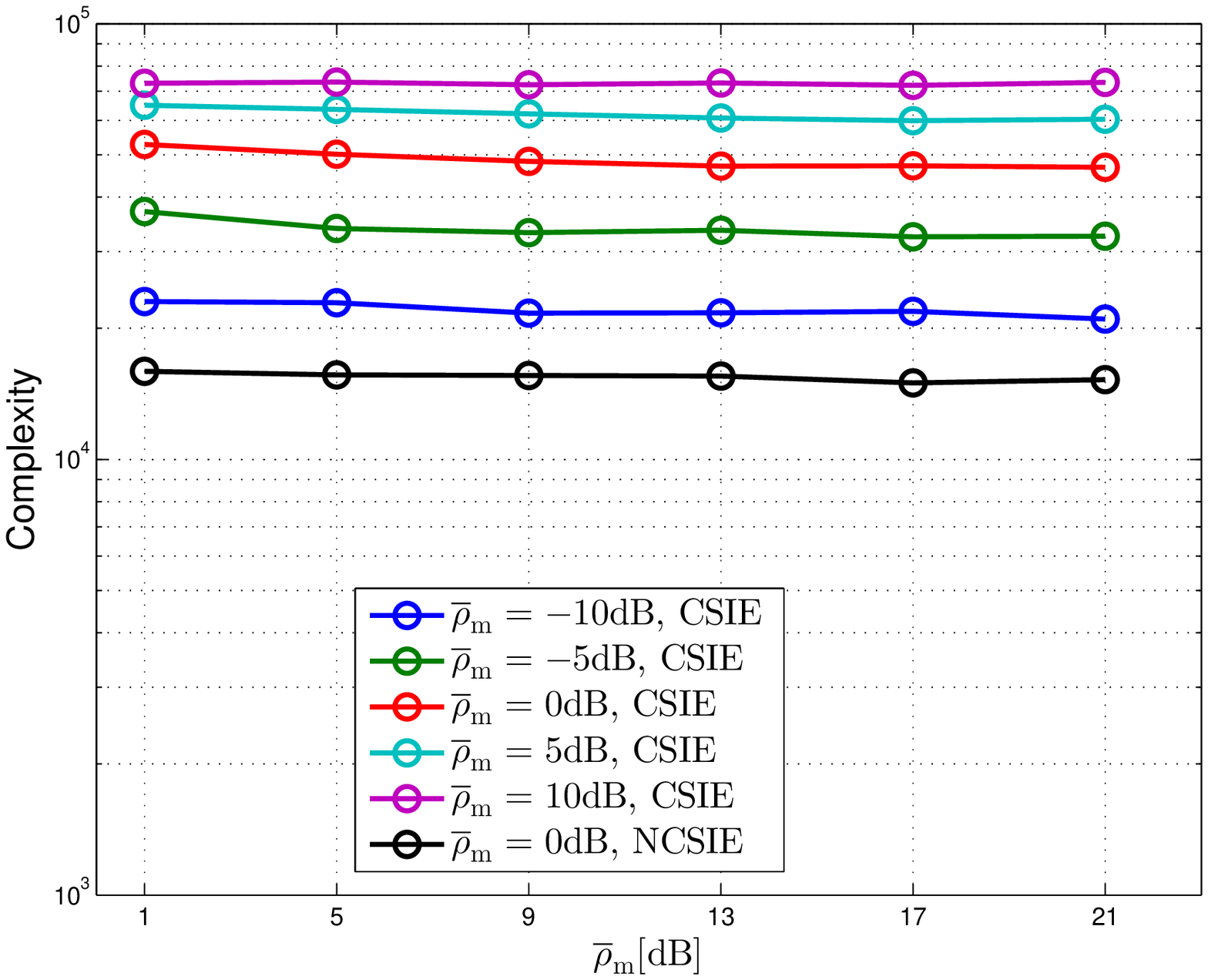} 
\caption{Complexity of BAB versus normalized SNR ${\overline{\rho}}_{\rm{m}}$ when $N_r=4$, $N_e=4$, $L=4$ and $N_{\rm{t}}=64$.}
\label{figure5}
\end{figure}

{\figurename} \ref{figure4} provides the complexity of the BAB, norm-based and ES method. As stated before, they can all be treated as tree search and the complexity of BAB is related with the number
of visited nodes. Therefore, it makes sense to use the number of visited nodes asking for updating operations during the tree search to measure the complexity of these algorithms \cite{b24}. As shown in {\figurename} \ref{figure4}, the norm-based method has the lowest complexity and the complexity of ES is much higher than that of the BAB. Based on above simulation results, the advantages of BAB are apparent for its optimality and low-complexity. To examine the robustness of the BAB method, {\figurename} \ref{figure5} shows the complexity versus $\overline{\rho}_{\rm{m}}$ and $\overline{\rho}_{\rm{e}}$. From this figure, it is clear that there are scarcely no great fluctuation in complexity for different $\overline{\rho}_{\rm{m}}$. Additionally, the complexity will increase as  $\overline{\rho}_{\rm{e}}$ rises up, which indicates that the BAB algorithm is more efficient under more secure transmission condition. Nevertheless, the fluctuation due to $\overline{\rho}_{\rm{e}}$ is not very large. Taken together, the proposed BAB method is robust and practical, especially for the scenario with NCSIE.    

\section{Conclusion}
\label{sction5}
This paper studies transmit antenna selection in massive MIMOME channels. An optimal algorithm based on branch-and bound search is proposed and discussed in the situations when the CSIE is available or unavailable. Simulation shows that branch-and-bound search can guarantee optimal performance with much lower complexity compared with exhaustive search. The proposed algorithm could serve as a benchmark in the future work on TAS algorithm design in massive MIMOME channels.

\vspace{12pt}

\begin{thebibliography}{00}
\bibitem{b1} A. Wyner, ``The wire-tap channel,'' {\emph{Bell Syst. Technol. J.}}, vol. 54, no. 8, pp. 1355--1387, 1975.
\bibitem{b2} C. E. Shannon, ``Communication theory of secrecy systems,'' {\emph{Bell Syst. Technol. J.}}, vol. 28, pp. 656--715, 1949.
\bibitem{b6} M. R. Islam, J. Kim and Md. S. Arefin, ``MIMOME channel secrecy capacity,'' in \emph{Proceedings of 11th International Conference on Computer and Information Technology (ICCIT 2008)}, 2008.
\bibitem{b4} A. Khisti and G. W. Wornell, ``Secure transmission with multiple antennas—part II: the MIMOME wire-tap channel,'' {\emph{IEEE Trans. Inf. Theory}}, vol. 56, no. 11, pp. 5515--5532, 2010.
\bibitem{b5} F. Oggier and B. Hassibi, ``The secrecy capacity of the MIMO wiretap channel,'' {\emph{IEEE Trans. Inf. Theory}}, vol. 57, no. 8, pp. 4961--4972, 2011.
\bibitem{b7} D. Kapetanovic, G. Zheng, and F. Rusek, ``Physical layer security for massive MIMO: An overview on passive eavesdropping and active attacks,'' {\emph{IEEE Commun. Magazine}}, vol. 53, no. 6, pp. 21--27, 2015.
\bibitem{b8} A. F. Molisch and M. Z. Win, ``MIMO systems with antenna selection,'' {\emph{IEEE Microwave Magazine}}, vol. 5, no. 1, pp. 46-56, 2004.
\bibitem{b12} N. Yang, P. L. Yeoh, M. Elkashlan, et al. ``Transmit antenna selection for security enhancement in MIMO wiretap channels,'' {\emph{IEEE Trans. Commun.}}, vol. 61, no. 1, pp. 144-154, 2013.
\bibitem{b13} N. Yang, H. A. Suraweera, I. B. Collings, and C. Yuen, ``Physical layer security of TAS/MRC with antenna correlation,'' {\emph{IEEE Trans. on Inf. Forensics and Security}}, vol. 8, no. 1, pp. 254--259, 2013.
\bibitem{b14} J. Zhu, Y. Zou, G. Wang, Y.-D. Yao, and G. K. Karagiannidis, ``On secrecy performance of antenna-selection-aided MIMO systems against eavesdropping,'' {\emph{IEEE Trans. on Vehicular Technology}}, vol. 65, no. 1, pp. 214--225, 2016.
\bibitem{b15} F. S. Al-Qahtani, Y. Huang, S. Hessien, et al. ``Secrecy analysis of MIMO wiretap channels with low-complexity receivers under imperfect channel estimation,'' {\emph{IEEE Trans. Inf. Forensics and Security}}, vol. 12, no. 2, pp. 257--270, 2017.
\bibitem{b17} H. Lei, M. Xu, I. S. Ansari, et al. ``On secure underlay MIMO cognitive radio networks with energy harvesting and transmit antenna selection,'' {\emph{IEEE Trans. Green Commun. and Networking}}, vol. 1, no. 2, pp. 192--203, 2017.
\bibitem{b18} S. Asaad, A. Bereyhi, R. R. Müller, and A. M. Rabiei, ``Optimal number of transmit antennas for secrecy enhancement in massive MIMOME channels,'' in \emph{IEEE Global Communications Conference (GLOBECOM)}, 2017. 
\bibitem{b20} A. Bereyhi, S. Asaad, R. R. Müller, et al. ``On Robustness of Massive MIMO Systems Against Passive Eavesdropping under Antenna Selection,'' to be presented in \emph{IEEE Global Communications Conference (GLOBECOM)}, 2018.
\bibitem{b22} G. J. Foschini and M. J. Gans, ``On limits of wireless communications in a fading environment when using multiple antennas,'' \emph{Wireless Pers. Commun.}, vol. 1, no. 2, pp. 41--49, 1996.
\bibitem{b23} M. P. Narendra, K. Fukunaga, ``A branch and bound algorithm for feature subset selection,'' {\emph{IEEE Trans. Computers}}, vol. C-26, no. 9, pp. 917--922, 1977.
\bibitem{b24} Y. Gao, H. Vinck, T. Kaiser, ``Massive MIMO antenna selection: Switching architectures, capacity bounds, and optimal antenna selection algorithms,'' {\emph{IEEE Trans. Signal Process.}}, vol. 66, no. 5, pp. 1346--1360, 2018.
\bibitem{b25} R. A. Horn, {\emph{Matrix Analysis}}. New York, NY, USA: Cambridge Univ. Press, 1986.
\end{thebibliography}
\end{document}